\documentclass[sigconf]{acmart}
\usepackage[nolist,nohyperlinks]{acronym}
\usepackage{multirow}
\usepackage{graphicx}
\usepackage{amsmath}

\AtBeginDocument{%
  }

\copyrightyear{2024}
\acmYear{2024}
\setcopyright{acmlicensed}\acmConference[RecSys '24]{18th ACM Conference
on Recommender Systems}{October 14--18, 2024}{Bari, Italy}
\acmBooktitle{18th ACM Conference on Recommender Systems (RecSys '24),
October 14--18, 2024, Bari, Italy}
\acmDOI{10.1145/3640457.3688139}
\acmISBN{979-8-4007-0505-2/24/10}

\begin{document}



\title{Transformers Meet ACT-R: Repeat-Aware and Sequential Listening Session Recommendation}
\author{Viet-Anh Tran}
\affiliation{
  \institution{Deezer Research}
  \city{}
  \country{}
}
\email{research@deezer.com}

\author{Guillaume Salha-Galvan}
\affiliation{
  \institution{Deezer Research}
  \city{}
  \country{}
}

\author{Bruno Sguerra}
\affiliation{
  \institution{Deezer Research}
  \city{}
  \country{}
}

\author{Romain Hennequin}
\affiliation{
  \institution{Deezer Research}
  \city{}
  \country{}
}

\renewcommand{\shortauthors}{Viet-Anh Tran, Guillaume Salha-Galvan, Bruno Sguerra, \& Romain Hennequin}

\begin{acronym}
    \acro{DL}{deep learning}
    \acro{NLP}{natural language processing}
    \acro{MF}{Matrix Factorization}
    \acro{HR}{Hit Rate}
    \acro{DCG}{Discounted Cumulative Gain}
    \acro{NDCG}{Normalized Discounted Cumulative Gain}
    \acro{CF}{Collaborative Filtering}
    \acro{RNN}{recurrent neural networks}
    \acro{GRU}{Gated Recurrent Unit}
    \acro{LSTM}{Long Short-Term Memory}
    \acro{CNN}{convolutional neural networks}
    \acro{GNN}{graph neural networks}
    \acro{ACT-R}{Adaptive Control of Thought—Rational}
    \acro{SR}{sequential recommendation}
    \acro{SVD}{singular value decomposition}
    \acro{NBR}{Next Basket Recommendation}
    \acro{RepR}{Repetition Ratio}
    \acro{BPR}{Bayesian Personalized Ranking}
\end{acronym}

\begin{abstract}
  Music streaming services often leverage sequential recommender systems to predict the best music to showcase to users based on past sequences of listening sessions. 
  Nonetheless, most sequential recommendation methods ignore or insufficiently account for repetitive behaviors. 
 This is a crucial limitation for music recommendation, as repeatedly listening to the same song over time is a common phenomenon that can even change the way users perceive this song.
  In this paper, we introduce PISA (\underline{P}sychology-\underline{I}nformed \underline{S}ession embedding using \underline{A}CT-R), a session-level sequential recommender system that overcomes this limitation. 
  PISA employs a Transformer architecture learning embedding representations of listening sessions and users using attention mechanisms inspired by Anderson's ACT-R (\underline{A}daptive
\underline{C}ontrol of \underline{T}hought-\underline{R}ational), a cognitive architecture modeling human information access and memory dynamics. This approach enables us to capture dynamic and repetitive patterns from user behaviors, allowing us to effectively predict the songs they will listen to in subsequent sessions, whether they are repeated or new ones.
We demonstrate the empirical relevance of PISA using both publicly available listening data from Last.fm and proprietary data from Deezer, a global music streaming service, confirming the critical importance of repetition modeling for sequential listening session  recommendation.
Along with this paper, we publicly release our proprietary dataset to foster future research in this field, as well as the source code of PISA to facilitate its future use.

\end{abstract}

\ccsdesc[300]{Information systems~Recommender systems}
\ccsdesc[300]{Information systems~Personalization}

\keywords{Sequential Recommendation, Repetition Modeling, Transformers, Music Streaming Service, Adaptive Control of Thought-Rational.}


\maketitle


\section{Introduction}
Recommender systems are essential to online platforms providing access to large catalogs, such as music streaming services~\cite{schedl2018current,jacobson2016music}. They mitigate information overload by identifying the most relevant content to showcase to users, e.g., personalized song selections for a music streaming service~\cite{pereira2019online}.
Moreover, recommender systems are regarded as effective tools to help users discover new content and improve their online experience \cite{schedl2018current,zhang_csur19}. Consequently, in recent years, researchers and practitioners have devoted significant efforts to develop better systems that would more accurately model user preferences on these services \cite{covington2016deep,gomez2015netflix,tran_recsys21,briand2021semi,mu2018survey}.

In particular, we have observed a growing interest in \textit{sequential} recommender systems \cite{hidasi_iclr16, trinh_recsys17, hu_sigir20, zhou_aaai18, zhang_aaai19, you_www19, kang_icdm18, li_wsdm20, fang_tois20,ren_aaai19,rappaz_recsys21,guo2022reinforcement,zhang2022enhancing}. 
Unlike static collaborative filtering  approaches \cite{hu_icdm08, rendle_auai09, schedl2018current}, these systems aim to capture the \textit{dynamic} dimension of user preferences \cite{moore_ismir13, liu_ijcai15}, e.g., musical tastes that would evolve over time \cite{quadrana2018modeling}. 
They typically predict the best content to recommend to users at a given time based on past observed sequences of user-content interactions~\cite{fang_tois20, wang_csur21,quadrana_csur21,tran_sigir23}, e.g., past listening sessions on a music streaming service \cite{pereira2019online, hansen_recsys20}.  To this end, they often build upon advances in deep learning techniques for sequence modeling, including recurrent neural networks \cite{rumelhart_nature86} and attention mechanisms~\cite{vaswani_nips17}.

Nonetheless, as further detailed in Section~\ref{s2}, most existing methods for sequential recommendation omit or insufficiently account for \textit{repetitive} patterns in interactions. We believe this to be a crucial limitation, especially for music-related applications~\cite{cheng_ijcai17, wang_inforet18, pereira_recsys19, hansen_recsys20, wang_multimedia21}, as repeatedly listening to the same music over time is rather frequent~\cite{gabbolini_recsys21,sguerra2022discovery}.
Repeated exposure also plays a key role in the music discovery process and changes the way users perceive songs~\cite{sguerra2022discovery}. 
To our knowledge, few studies have explored this important aspect for sequential music recommendation \cite{reiterhaas_recsys21, marta_recsys23}. They have characterized repetitive behavior using \ac{ACT-R} \cite{anderson_psyreview04, bothell20}, a cognitive architecture that encompasses a module modeling the dynamics of human memory access. However, their use of ACT-R was limited to inference only, not extending to model training. Also, they did not capture the \text{dynamic} dimension of preferences, a piece of crucial information for sequential recommendation~\cite{kang_icdm18,sun_cikm19,tran_sigir23}. Therefore, the challenge of sequentially recommending music while accurately modeling \textit{dynamic and repetitive} patterns from past listening actions remains relatively~open.

In this paper, we propose to tackle this important challenge. We focus on the \textit{sequential
 listening session recommendation} setting, i.e., on predicting the songs users will listen to during their next listening session based on a sequence of past sessions.
We show that, for this problem, repeat-aware and dynamic sequential recommendation is not only feasible but also exhibits promising performance. Our approach uniquely combines  advances in Transformer-based sequence modeling with psychology-based repetitive behavior modeling using ACT-R, all within a cohesive framework. More specifically, our contributions in this paper are~listed~as~follows:

\begin{itemize}
    \item We introduce PISA (Psychology-Informed Session embedding using ACT-R), a Transformer system for repeat-aware and sequential listening session recommendation. PISA leverages attention mechanisms inspired by ACT-R components to learn embedding representations of sessions and users accounting for sequential and repetitive patterns from past listening actions. These representations enable PISA to accurately predict the songs users will engage with in future sessions, whether they are repeated or new ones.
    \item  We demonstrate the empirical effectiveness of our approach, through comprehensive 
    experimental validation on public listening data from the music website Last.fm and proprietary data from the global music streaming service Deezer. Our results confirm the importance of modeling repetitive patterns for sequential listening session recommendation.
    \item We release the source code of PISA to ensure the reproducibility of our results and facilitate its future use. Additionally, we release our Deezer dataset of listening sessions. We hope that making these industrial resources available will benefit the scientific community and foster research in the domain. 
\end{itemize}

The remainder of this paper is organized as follows. In Section~\ref{s2}, we introduce our problem more formally and review the relevant related work. In Section~\ref{section_pepsi}, we present PISA. We report and discuss our experimental results in Section~\ref{experiment}, and conclude in Section~\ref{conclusion}.

\section{Preliminaries}
\label{s2}

We begin this section with a formal definition of the problem we aim to solve in this paper. We subsequently review the related work.

\begin{figure*}[ht!]
  \centering
  \includegraphics[width=0.88\textwidth]{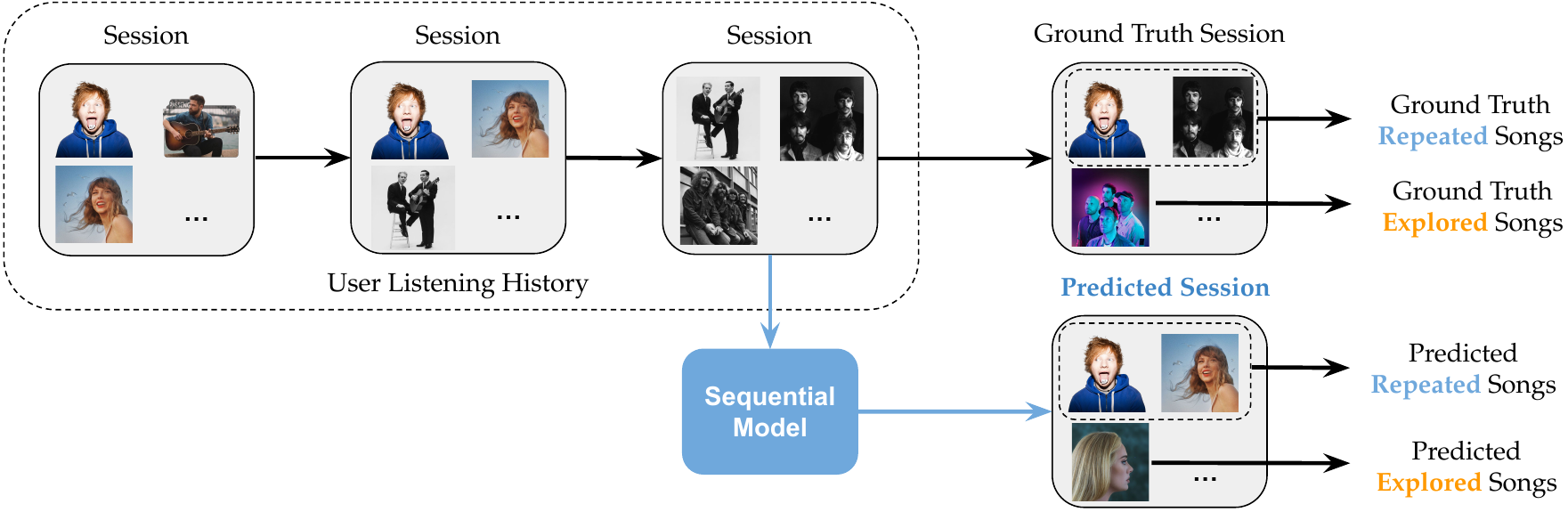} 
  \caption{Illustration of the  listening session recommendation problem. Based on a past sequence of sessions made by a user on a streaming service, we aim to predict the set of songs this user will listen to during their next session. The user may exhibit \textit{repetitive} behaviors, i.e., relistening to songs from previous sessions, as well as \textit{explorative} behaviors, i.e., listening to new songs.}
  \label{fig:prob}
\end{figure*}

\subsection{Problem Formulation}
\label{prob_formulation}

\subsubsection{Setting and Objective} 
Unlike movies or books, songs are short media pieces, with comparatively lighter engagement and most frequently listened to in succession on music streaming services like Spotify or Deezer~\cite{schedl2018current, hansen_recsys20}. In this paper, we use the term \textit{listening session} to refer to a set of songs listened to within a specific time frame, according to criteria established by such services.

Our objective is to build a recommender system that, based on past observed \textit{sequences of successive listening sessions} from users, accurately predicts the next songs these same users will listen to in their subsequent sessions. 
We note that users' musical listening habits are complex, with reported \textit{dynamic} patterns, such as evolving preferences over time~\cite{hansen_recsys20,tran_sigir23,sanna2021next}. Besides, repeatedly listening to the same songs on music streaming services is rather frequent \cite{conrad2019extreme,reiterhaas_recsys21,tsukuda2020explainable}. Hence, sequences of listening sessions also reflect \textit{repetitive} patterns. In particular, recommending the same song again in a future session may be a relevant option. This aspect contrasts with other application domains such as movie recommendation, where such repetitions are often unwelcome~\cite{reiterhaas_recsys21,schedl2018current}.

\subsubsection{Mathematical Formalization} More formally, we consider in this paper a set~$\mathcal{U}$ of users on Deezer, and a set~$\mathcal{V}$ of songs available on this same service. 
We assume that, for each user $u~\in~\mathcal{U}$, we have observed\footnote{In practice, some users may have engaged in more than $L$ sessions. For these users, one can consider a subset of $L$ sessions, e.g., the $L$ most recent ones.} $L \in \mathbb{N}^*$ past listening sessions on the service. We denote by
$S^{(u)} = (s^{(u)}_1, s^{(u)}_2, \dots, s^{(u)}_{L})$
the ordered sequence of $L$ listening sessions made by the user~$u$. Each element $s^{(u)}_l \in S^{(u)}$,  with $l \in \{1,\dots,L\}$, corresponds to the $l$-th listening session of the user $u$ on the service. It materializes as a set of $K \in \mathbb{N}^*$ songs\footnote{Following the approach of Hansen et al. \cite{hansen_recsys20}, we only consider the first $K$ songs of each session. This decision is based on the observation that, in longer sessions, the relevance of songs at the end to those at the beginning becomes increasingly uncertain.} listened to by the user $u$ during this session:
\begin{equation}
s^{(u)}_l = \{v^{(u)}_{l,1}, v^{(u)}_{l,2},...,v^{(u)}_{l,K}\},
\end{equation}
with $v^{(u)}_{l,k} \in \mathcal{V}, \forall k \in \{1,\dots,K\}.$ 
As $s^{(u)}_l$ is a set and not a sequence, we do not account for the order in which the $K$ songs of the same session are played. In this work, we treat them as an unordered collection. We focus on song inclusion in each session, while modeling the dynamics in sequences of successive sessions.

Using this formalism, the listening session recommendation problem under consideration in this paper consists in predicting:
\begin{equation}
s^{(u)}_{L+1} = \{v^{(u)}_{L+1, 1}, v^{(u)}_{L+1, 2},...,v^{(u)}_{L+1, K}\},
\end{equation}
i.e., the set of $K$ songs that $u$ will interact with in their next session $s^{(u)}_{L+1}$, based on $S^{(u)}$. Figure~\ref{fig:prob} illustrates this problem.
For model evaluation, we will compare the $K$ songs  predicted by our system for $s^{(u)}_{L+1}$ against the ``ground truth'' songs actually listened to by $u$, which could be new ones as well as repeats from previous sessions.



\subsection{Related Work}
\label{rel_work}
\subsubsection{Sequential and Next Basket Recommendation}
\label{nbr_biblios}
In recent years, there has been a growing interest in addressing sequential recommendation problems \cite{hidasi_iclr16, trinh_recsys17, hu_sigir20, zhou_aaai18, zhang_aaai19, you_www19, kang_icdm18, li_wsdm20, fang_tois20,ren_aaai19,guo2022reinforcement,zhang2022enhancing,tran_sigir23}. Two of the most popular sequential recommender systems are SASRec and BERT4Rec. SASRec~\cite{kang_icdm18} was the first to apply self-attention to identify relevant recommendable items from user-item interaction sequences. BERT4Rec~\cite{sun_cikm19} then employed bidirectional self-attention techniques for this purpose. These systems have been successfully used for various applications, including sequential music recommendation in diverse settings~\cite{tran_sigir23,moor_cikm23,pereira_recsys19,schedl2018current,bendada2023scalable}.

In particular, existing research often refers to the specific task of Section~\ref{prob_formulation} as \ac{NBR} \cite{li_infosys23,hu_sigir20,wan_cikm18,ariannezhad_sigir22}. The term ``basket'' designates a set of items consumed together, and the goal is to predict the next basket users will interact with, based on past basket sequences. Recent studies on \ac{NBR} primarily focused on capturing user preferences through transitions between baskets. This was often achieved using neural network approaches tailored for sequence modeling, such as \ac{RNN} \cite{yu_sigir16, bai_sigir18, hu_sigkdd19, le_ijcai19}, attention mechanisms \cite{sun_sigir20, yu_sigkdd20, chen_arxiv21, li_recsys23}, or using denoising techniques via contrastive learning  \cite{qin_sigir21}. 
In the music domain, the work of Hansen et al. \cite{hansen_recsys20} constitutes a noticeable effort to reframe listening session recommendation as an \ac{NBR} task. 
Authors proposed CoSeRNN, a RNN system learning context-dependent embedding representations of Spotify users and sessions in a common vector space, permitting to identify the most relevant sessions to sequentially recommend to these users depending on the context.

However, although baskets are key inputs in the cited works, their representation learning has received limited attention. Most methods used simple item aggregations like average and max pooling to form basket representations~\cite{hansen_recsys20, yu_sigir16, hu_sigkdd19, qin_sigir21, shen_tkdd22}. 
For example, Hansen et al. \cite{hansen_recsys20} averaged embeddings of songs from each session to represent sessions. Yu et al.~\cite{yu_ijcai23} noted that, through these operations, important information on baskets is lost, notably repetitive patterns which are crucial for repeat-aware modeling and recommendation.

\subsubsection{Modeling Repetitive Behaviors for NBR}
\label{repeatNBRbaselines}
The importance of repetition modeling for \ac{NBR} has been firmly established in recent research \cite{benson_sigkdd18, wan_cikm18, hu_sigkdd19, hu_sigir20, faggioli_umap20, ariannezhad_sigir22, li_infosys23, yu_ijcai23, ariannezhad_cikm21}. Benson et al.~\cite{benson_sigkdd18} were among the first ones to uncover repetitive patterns in sequences of sets. However, the stochastic model they developed to model these patterns operates under the restrictive assumption that the next set is composed solely of elements from previous sets. 
Hu et al. \cite{hu_sigkdd19} subsequently proposed Sets2Sets, an attention-based encoder-decoder framework overcoming this restriction, with application to medical and e-commerce NBR.
Yu et al. \cite{yu_sigkdd20} also introduced DNNTSP, a graph neural network predicting temporal sets including repeated and new items. Hu et al. \cite{hu_sigir20} argued that previous studies concentrated on the personalized item frequency information for individual users but overlooked the collaborative aspects of repetitive patterns among several users. They proposed TIFU-KNN, a nearest neighbor \ac{NBR} model surmounting this limitation with promising performance on repeated purchase modeling in grocery shopping. Concurrently, Faggioli et al. \cite{faggioli_umap20} introduced UP-CF@r, a recency-aware collaborative filtering model for \ac{NBR}. 
 Ariannezhad et al. \cite{ariannezhad_sigir22} studied ReCANet, an LSTM system for the sequential modeling of repeated consumption for each item. 
More recently, Yu et al. \cite{yu_ijcai23} introduced BRL, which aims to capture intra-basket correlations by applying hypergraph convolutions \cite{feng_aaai19} to each basket.

\subsubsection{Modeling Repetitive Behaviors for Music NBR}
\label{musicnbr}

The studies referenced in the preceding subsection mainly put the emphasis on e-commerce data and product repurchase applications. To our knowledge, there have been few investigations into repeat-aware NBR for sequential music recommendation. However, as explained in Section~\ref{prob_formulation}, repeatedly listening to the same songs over time is frequent \cite{conrad2019extreme,reiterhaas_recsys21,tsukuda2020explainable}.
Furthermore, repetitive behaviors are crucial in the music discovery process. In particular, repeated exposure to a song can significantly alter a user's perception and interest in that song, thereby affecting its relevance when being recommended~\cite{sguerra2022discovery}.

The few studies that did focus on relistening modeling for sequential music recommendation purposes have characterized repetitive
behaviors using Anderson's \ac{ACT-R}~\cite{anderson_psyreview04, bothell20}. Here we note that, beyond the scope of our work, several studies have also made use of ACT-R's modules for applications, including hashtag reuse modeling, mobile app usage prediction, job recommendation, and music genre preference modeling~\cite{lex2019impact,lex2020modeling,lacic2017beyond,lacic2019should,zhao2014context}. This well-established cognitive architecture, which we will also use in PISA, describes different human cognitive functions, particularly, encompassing a module modeling memory access dynamics.
Reiter-Haas et al. \cite{reiterhaas_recsys21} used the ACT-R memory module to predict music relistening behaviors in user sessions, showing superior prediction accuracy over baselines that select recent songs. However, Moscati et al. \cite{marta_recsys23} pointed out that their approach only recommends repeated songs that users have already listened to. 
 To overcome this limitation and also suggest novel songs, Moscati et al.~\cite{marta_recsys23} studied the combination of ACT-R with various collaborative filtering models, such as a \ac{BPR}~\cite{rendle_auai09}. They proposed an explainable two-stage scheme that initially involves pre-training a collaborative filtering model, followed by modifying recommendation scores using ACT-R during inference. 
 
We argue that previous approaches, however, suffer from  drawbacks. Firstly, their application of ACT-R was limited to inference only, not extending to model training. Secondly, they did not propose ways to integrate ACT-R with sequential recommender systems. The pure collaborative filtering models they considered did not capture the dynamic dimension of user preferences, embedded within user-song interaction sequences. However, as we will confirm in our experiments, capturing these dynamic patterns is essential for making effective predictions. 
Consequently, the challenge of sequentially recommending music while accurately reflecting the \textit{dynamic and repetitive} patterns from past actions remains relatively~open.

\section{Sequential and Repeat-Aware
Listening Session Recommendation}
\label{section_pepsi}

In this section, we introduce our PISA system, which aims to address the challenge outlined at the end of Section~\ref{s2} by combining:
\begin{itemize}
    \item  A dynamic modeling of listening session sequences, using a Transformer-based neural network architecture;
    \item A psychology-informed modeling of relistening behaviors across sessions, using the ACT-R framework.
\end{itemize}
We begin by presenting the ACT-R declarative module employed in PISA in Section~\ref{actr_framework}. Then, we detail how PISA learns embedding representations of listening sessions in Section~\ref{sess_emb}, represents users in Section~\ref{sess-user}, performs session recommendation using these representations in Section~\ref{prediction}, and our training procedure in Section~\ref{training}.

\subsection{ACT-R Framework}
\label{actr_framework}
Building upon prior studies on music recommendation \cite{reiterhaas_recsys21, marta_recsys23, sguerra2023ex2vec}, we consider the declarative module of the \ac{ACT-R} framework \cite{anderson_psyreview04} to model relistening behaviors. This module is responsible for modeling the dynamics of information activation and forgetting of the human memory. Its use in music consumption prediction is based on the findings that music preferences relate to memory and exposure~\cite{szpunar2004liking, peretz1998exposure}, i.e., the more a user is exposed to a song, the better (to some extent) the user understands it. This premise is reinforced by data analysis from music streaming services~\cite{sguerra2022discovery, sguerra2023ex2vec}. ACT-R's declarative module includes a series of activation functions that model the way the human mind accesses stored information and has been rather successful in modeling repetitive behaviors~\cite{lex2019impact}.  
Precisely, to model the ease with which a user $u \in \mathcal{U}$ would retrieve a song $v \in \mathcal{V}$ from memory, this module would sum values of various components, each reflecting a specific aspect of how the mind stores and accesses information~\cite{bothell20}:
\begin{itemize}
    \item \text{Base-level component} $\text{BL}^{(u)}_v$: reflects the observation that information activated more frequently or recently is more easily retrieved from memory. For music preferences, songs with a high base-level component would be ``hot tracks'' that the user has been listening to with high frequency or recently.

    
    \item \text{Spreading component} $\text{SPR}^{(u)}_v$: favors songs that are often co-listened to in the same context, in our case, in listening sessions. It operates on the principle that if a song 
$v$ frequently appears in sessions alongside certain other songs, then the presence of these songs in a given session will enhance the memory activation of $v$ during that session.
    
    \item \text{Partial matching component} $\text{P}^{(u)}_v$: further enables the activation of similar songs, based on their musical characteristics. For instance, if $v$ is a rock song, the presence of a closely related rock song $v'$ (from a content perspective) in the session would increase the memory activation of $v$, even if listening data do not show them as frequently co-occurring. 
\end{itemize}
The declarative module may include other components, notably as a noise term accounting for randomness in user behavior. However, due to the subpar performance of noise-inclusive models in past studies~\cite{reiterhaas_recsys21}, we focus on the above three components within PISA.



\begin{figure*}[t]
  \centering
  \includegraphics[width=0.88\textwidth]{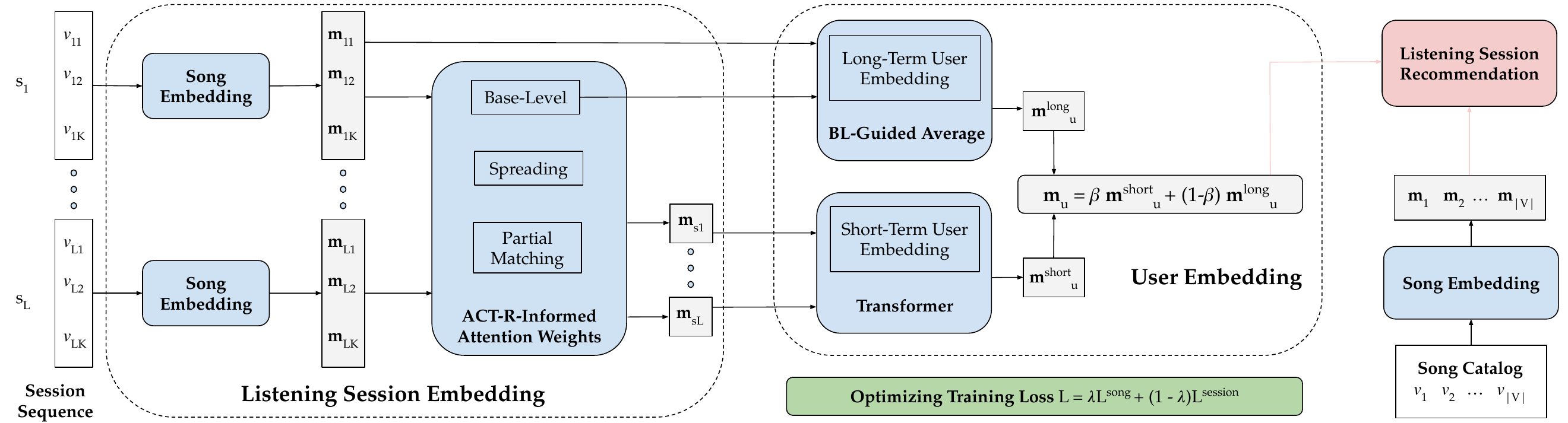} 
  \caption{Architecture of the PISA system presented in Section~\ref{section_pepsi} for repeat-aware sequential listening session recommendation.}
  \label{fig:pepsi}
\end{figure*}

\subsection{Session Embedding}
\label{sess_emb}

\subsubsection{Overview} PISA learns session embedding representations using attention weights guided by ACT-R components. We denote by $\mathbf{M} \in \mathbb{R}^{|\mathcal{V}| \times d}$ a song embedding matrix, in which rows are embedding vectors $\mathbf{m}_v \in \mathbb{R}^d$ representing each song $v \in \mathcal{V}$, for some dimension $d \in \mathbb{N}^*$. This matrix can be pre-computed (using content-based or collaborative filtering methods \cite{koren2015advances}) or directly learned within PISA (see Section~\ref{training}). Given these song embedding vectors, PISA represents the session $s^{(u)}$ of some user $u \in \mathcal{U}$ by a session embedding vector $\mathbf{m}_{s^{(u)}} \in \mathbb{R}^d$ in the same space, as follows:
\begin{equation}
\label{sess_rep}
    \mathbf{m}_{s^{(u)}} = \sum_{v \in s^{(u)}} w_v \mathbf{m}_{v}.
\end{equation}
The terms $w_v \geq 0$, with $\sum_{v \in s^{(u)}} w_v = 1$, are ACT-R-informed attention weights associated with each song in the session, with:
\begin{equation}
    w_v = w_{\text{BL}}\text{BL}^{(u)}_{v} + w_{\text{SPR}} \text{SPR}^{(u)}_{v} + w_{\text{P}} \text{P}^{(u)}_{v}.
\end{equation}
The remainder of Section~\ref{sess_emb} details how we compute the $\text{BL}^{(u)}_{v}$, $\text{SPR}^{(u)}_{v}$, and $\text{P}^{(u)}_{v}$ components. Besides, $w_{\text{BL}}, w_{\text{SPR}}$, and $w_{\text{P}}$ are global parameters learned using a one-layer feedforward neural network processing these components with shared weights~across~users.

\subsubsection{Base-Level Component} In line with prior work~\cite{reiterhaas_recsys21,marta_recsys23},~we~set: 
\begin{equation}
\label{bll_eq}
    \text{BL}^{(u)}_v = \text{softmax}_{s^{(u)}}\Big(\sum_{k}(t_{\text{ref}} - t^{(uv)}_{k})^{-\alpha}\Big).
\end{equation}
$t_{\text{ref}}$ denotes the reference or prediction time, and $t^{(uv)}_{k}$ indicates the time of the $k$-th listening of song $v$ by user $u$ ($t^{(uv)}_{k} < t_{\text{ref}}$). The parameter $\alpha \in \mathbb{R}^+$ serves as a time decay factor modeling the forgetting of past listens. The softmax operation normalizes values across all songs from the same session (thus, $\sum_{v\in s^{(u)}}\text{BL}^{(u)}_v = 1$). In essence, $\text{BL}^{(u)}_v$ increases with the frequency and recency of the occurrences of $v$ within listening sessions of $u$.
Consequently, songs that are played often and recently by $u$ 
 will carry more weight in the way PISA will represent their sessions. This, in turn, will affect how PISA represents $u$ (see Section~\ref{sess-user}) and, ultimately, which songs will be recommended to $u$ in future listening sessions (see Section~\ref{prediction}).

\subsubsection{Spreading Component}
Regarding $\text{SPR}^{(u)}_v$, we begin by constructing the song co-occurrence matrix $\textbf{F} \in \mathbb{R}^{|V| \times |V|}$~\cite{le_ijcai19}. Each element $\textbf{F}_{ij}$ represents the number of times songs $i$ and $j$ have appeared together in the same session, for all $i \neq j$, across all sessions used for model training. 
Then, we compute the song correlation matrix $\textbf{C} = \textbf{D}^{-\frac{1}{2}} \textbf{F} \textbf{D}^{-\frac{1}{2}},$
where $\textbf{D}$ is the diagonal matrix verifying $\textbf{D}_{ii} = \sum_j \textbf{C}_{ij}$ for all $i$ and $\textbf{D}_{ij} = 0$ for all $i \neq j$. 
Finally, we compute the spreading activation for each song $v$ in a session $s^{(u)}$ as follows:
\begin{equation}
    \text{SPR}^{(u)}_v = \sum_{v' \in s^{(u)}, v' \neq v} \textbf{C}_{vv'}.
    \label{spr}
\end{equation}
$\text{SPR}^{(u)}_v$ increases when songs close to $v$ according to $\textbf{C}$ appear in the session.
From the ACT-R perspective, these correlated songs enhance the memory activation for $v$, thereby giving it more weight in the session representation. Again, this will affect how PISA represents $u$ (Section~\ref{sess-user}) and provides recommendations (Section~\ref{prediction}).

\subsubsection{Partial Matching Component}
We aim to account for the effect of songs correlated with $v$ and appearing in the session, thereby increasing $v$'s memory activation, but through other means than co-listening patterns from Equation~\eqref{spr}.
To this end, we compute dot products of song embedding vectors to measure similarities:
\begin{equation}
    \text{P}^{(u)}_v = \sum_{v' \in s^{(u)}, v' \neq v} \mathbf{m}^{\intercal}_v \mathbf{m}_{v'}.
\end{equation}
This term can encompass complementary information with respect to $\text{SPR}^{(u)}_v$, such as content-based similarities. In this case, the presence of songs that are musically akin to 
$v$ in the same session would enhance the memory activation for $v$, even though this musical similarity is not reflected in the co-listening patterns of $\textbf{C}$.

\subsection{User Embedding}
\label{sess-user}

\subsubsection{Overview}
We now explain how PISA learns user embedding representations summarizing their preferences, in the same embedding space as songs and sessions.
We follow the prevalent approach in sequential recommendation~\cite{fang_tois20,adomavicius2011context,hansen_recsys20,tran_sigir23} where each user $u \in \mathcal{U}$ is represented by a vector $\mathbf{m}_u \in \mathbb{R}^d$ defined as the combination of:
\begin{itemize}
    \item a ``long-term'' embedding vector $\mathbf{m}^{\text{long}}_u \in \mathbb{R}^d$, which captures intrinsic user preferences, independent of the context;
    \item a ``short-term'' embedding vector $\mathbf{m}^{\text{short}}_u \in \mathbb{R}^d$, which reflects how recent sessions would affect user preferences and the perception of recommendations at a given time. In this work, we leverage a Transformer architecture to dynamically model sequences of past listening sessions for each user. \end{itemize}
PISA fuses short-term and long-term user preferences as follows:
\begin{equation}
    \mathbf{m}_u = \beta \mathbf{m}^{\text{short}}_u + (1 - \beta) \mathbf{m}^{\text{long}}_u.
    \label{eq-user-embedding}
\end{equation}
We learn the parameter $\beta \in [0, 1]$ using a one-layer feedforward neural network processing the concatenated vector $[\mathbf{m}^{\text{short}}_u; \mathbf{m}^{\text{long}}_u]$. 
The remainder of Section~\ref{sess-user} explains how we learn $\mathbf{m}^{\text{short}}_u$ and~$\mathbf{m}^{\text{long}}_u$.

\subsubsection{Long-Term Representation}
Users have diverse musical tastes. Recommender systems often represent ``long-term'' preferences using a weighted average of song embedding vectors from their historical listening data \cite{jing_recsys20,wu_recsys20,hansen_recsys20}. In PISA, we integrate weights from ACT-R's base-level activation in this averaging process:

\begin{equation}
    \mathbf{m}^{\text{long}}_u = \sum_{v \in \text{Top-BL}^{(u)}} \text{BL}^u_{v} \mathbf{m}_{v},
\end{equation}
where the set $\text{Top-BL}^{(u)}$ comprises the 20 songs listened to by $u$ in their previous sessions having the highest $\text{BL}^{(u)}_{v}$ activation values, as computed in Equation~\eqref{bll_eq} and normalized using a softmax function.
By focusing on frequently repeated songs, we aim to indirectly denoise listening history data, helping PISA concentrate on the songs that most accurately reflect each user's musical tastes.

\subsubsection{Short-Term Representation}
In practice, past interactions can significantly alter user preferences and perceptions of each recommended song~\cite{adomavicius2011context,fang_tois20}.
Consequently, the most relevant songs to recommend to $u$ at time $T+1$ might not be those immediately adjacent to $\textbf{m}^{\text{long}}_u$ in the embedding space, but rather those near the translated representation $\beta \mathbf{m}^{\text{short}}_u + (1 - \beta) \mathbf{m}^{\text{long}}_u$. 

In PISA, we leverage self-attention mechanisms \cite{vaswani_nips17, kang_icdm18} to accurately capture the dynamics of session sequences and learn $\textbf{m}^{\text{short}}_u$. Firstly, we represent each sequence $S^{(u)} = (s^{(u)}_1, \dots, s^{(u)}_{L})$ by the matrix
$\mathbf{E}_{S^{(u)}} = [ \mathbf{m}_{s^{(u)}_1}, \mathbf{m}_{s^{(u)}_2},\dots, \mathbf{m}_{s^{(u)}_L} ]^{\intercal} \in \mathbb{R}^{L \times d}$,
where the representation for each session $\mathbf{m}_{s^{(u)}_l} \in \mathbb{R}^d$, with $l \in \{1, \dots, L\}$, is obtained using Equation~\eqref{sess_rep}. Secondly, to take into account the influence of each session's position within the sequence $S^{(u)}$, we enrich $\mathbf{E}_{S^{(u)}}$ with learnable positional embedding vectors $\mathbf{P} = [ \mathbf{p}_1, \dots, \mathbf{p}_L ]^{\intercal} \in \mathbb{R}^{L \times d}$, obtaining our final \text{input matrix}:
\begin{equation}
  \mathbf{X}^{(0)} = [ \mathbf{x}^{(0)}_1, \dots, \mathbf{x}^{(0)}_L]^{\intercal} \in \mathbb{R}^{L \times d},
\end{equation} where $\mathbf{x}^{(0)}_l = \mathbf{m}_{s^{(u)}_l} + \mathbf{p}_l, \forall l \in \{1,\dots, L\}$.
Thirdly, we pass $\mathbf{X}^{(0)}$ through $B \in \mathbb{N}^*$ stacked \text{self-attention blocks} (SABs). The output of the $b^{\text{th}}$ block is $\mathbf{X}^{(b)} = \text{SAB}^{(b)}(\mathbf{X}^{(b-1)})$, for $b \in \{1, \dots, B\}$. The SAB contains a \text{self-attention layer} $\text{SAL}(\cdot)$ with $H \in \mathbb{N}^*$ heads, followed by a \text{feedforward~layer}~$\text{FFL}(\cdot)$:
\begin{align}
    \text{SAL}(\mathbf{X}) &= \text{MultiHead}(\{\mathbf{X}^{\text{Att}}_j\}^H_{j=1}) = \text{Concat}(\mathbf{X}^{\text{Att}}_1,\dots,\mathbf{X}^{\text{Att}}_H) \mathbf{W}^O, \nonumber \\
    \text{SAB}(\mathbf{X}) &= \text{FFL}(\text{SAL}(\mathbf{X})) = \text{ReLU}(\mathbf{X}^{\text{Att}} \mathbf{W}_1 + \mathbf{b}_1) \mathbf{W}_2 + \mathbf{b}_2,
\end{align}
where $\mathbf{X} \in \mathbb{R}^{L \times d}$ is the input of each block, $\mathbf{W}^O \in \mathbb{R}^{d \times d}$ is the projection matrix for the output, and $\mathbf{W}_1$, $\mathbf{W}_2 \in \mathbb{R}^{d \times d}$ and $\mathbf{b}_1, \mathbf{b}_2 \in \mathbb{R}^{1 \times d}$ are weights and biases for the two layers of the FFL network. $\mathbf{X}^{\text{Att}}_j = \text{softmax}(\mathbf{A}_j / \sqrt{d}) \mathbf{V}_j$ is the output of the head $j$, where $\mathbf{A}_j = \mathbf{Q}_j \mathbf{K}^{\intercal}_j \in \mathbb{R}^{L \times L}$ with $\mathbf{Q}_j = \mathbf{X} \mathbf{W}^{(j)}_Q, \mathbf{K}_j = \mathbf{X} \mathbf{W}^{(j)}_K$ and $\mathbf{V}_j = \mathbf{X} \mathbf{W}^{(j)}_V$. $\mathbf{W}^{(j)}_Q, \mathbf{W}^{(j)}_K, \mathbf{W}^{(j)}_V \in \mathbb{R}^{d \times \frac{d}{H}}$ are learnable parameters. The final short-term embedding vector $\mathbf{m}^{\text{short}}_u$ corresponds to the final model output at position $L$ after $B$ attention blocks:
\begin{equation}
\mathbf{m}^{\text{short}}_u = \mathbf{X}_L^{(B)}.
\end{equation}
We emphasize that, as PISA processes session embeddings from Section~\ref{sess_emb} at this stage, ACT-R components influence the short-term deviation $\mathbf{m}^{\text{short}}_u$ and, consequently, the final user embedding vector $\mathbf{m}_u$. Songs with the highest ACT-R activation levels have the greatest impact on user representations in the embedding space, influencing final recommendations in the next Section~\ref{prediction}.

\subsection{Listening Session Recommendation}
\label{prediction}
At this stage, PISA has mapped songs, sessions, and users in the same embedding space, which facilitates similarity comparisons. As outlined in Section~\ref{prob_formulation}, our ultimate goal is to predict the set of songs that each user $u$ will listen to in their next session, after $S^{(u)}$. For this purpose, we adopt a latent factor approach \cite{kang_icdm18} in PISA, scoring the relevance of each song $v \in \mathcal{V}$ by evaluating $r_{L+1}(v) = \mathbf{m}_{u}^{\intercal} \mathbf{m}_v \in \mathbb{R}.$
As we consider sessions of length $K$, PISA recommends the top-$K$ songs having the highest relevance for~each~user.

\subsection{Training Procedure}
\label{training}

PISA requires optimizing multiple weights, covering those within our Transformer and feedforward neural networks, in addition to positional vectors. Moreover, if a pre-computed matrix $\textbf{M}$ is unavailable, PISA would directly learn embedding vectors $\mathbf{m}_v$ for each $v \in \mathcal{V}$. We denote by $\Theta$ the entire set of weights to optimize.
For this purpose, we consider a training set $\mathcal{S}$ of session sequences. For each sequence $S^{(u)} = (s^{(u)}_1, s^{(u)}_2, \dots, s^{(u)}_{L}) \in \mathcal{S}$, we create sub-sequences comprising the first $l$ sessions only, for $l \in \{1,\dots, L-1\}$. 
Also, we use $\mathbf{m}_{u,l}$ to denote the user embedding vector of $u \in \mathcal{U}$ that PISA would compute by processing, not the entire sequence $S^{(u)}$ as in Equation~\eqref{eq-user-embedding}, but only the first $l$ sessions of this sequence.
When recommending to $u$ a set of $K$ songs to extend these sub-sequences, we expect PISA to assign high relevance scores to songs in $s^{(u)}_{l+1}$, i.e., the ``ground truth'' $K$ songs listened to by $u$ after $l$ sessions. Simultaneously, we expect PISA to assign lower scores to songs from $o^{(u)}_{l+1}$, a randomly sampled ``negative'' set of $K$ songs from $\mathcal{V} \setminus s^{(u)}_{l+1}$.

To this end, we optimize $\Theta$ through the gradient descent minimization of the loss 
$\mathcal{L}(\Theta) = \lambda \mathcal{L}^{\text{song}}(\Theta) + (1 - \lambda) \mathcal{L}^{\text{session}}(\Theta),$
 where $\lambda \in [0, 1]$ is an hyperparameter to set, and where:
 \begin{equation}
    \mathcal{L}^{\text{song}}(\Theta) = \sum_{S^{(u)} \in \mathcal{S}} \sum^{L-1}_{l=1} \sum_{v \in s^{(u)}_{l+1}, v' \in o^{(u)}_{l+1}} \ln\bigl(1 + e^{-(\mathbf{m}^{\intercal}_{u,l}\mathbf{m}_v - \mathbf{m}^{\intercal}_{u,l}\mathbf{m}_{v'})}\bigr),
    \label{loss}
\end{equation}
\begin{equation}
    \mathcal{L}^{\text{session}}(\Theta) = \sum_{S^{(u)} \in \mathcal{S}} \sum^{L-1}_{l=1} \bigl(1 - \mathbf{m}^{\intercal}_{u,l}\mathbf{m}_{s^{(u)}_{l+1}}\bigr).
\end{equation}
The first term, $\mathcal{L}^{\text{song}}(\Theta)$, is a \ac{BPR} loss \cite{rendle_auai09}. We employ it to ensure that dot products between user embedding vectors after $l$ sessions and embedding vectors of songs from their $(l+1)$-th session are higher than those with songs they did not listen to.
The second term, $\mathcal{L}^{\text{session}}(\Theta)$, corresponds to session-level loss of Hansen et al.~\cite{hansen_recsys20}. We add it to further ensure that such user embedding vectors have high dot products with their respective $(l+1)$-th session.

\section{Experimental Analysis}
\label{experiment}
We now evaluate our PISA system. We present our experimental setting in Sections~\ref{datasets} to \ref{implem}, and discuss our results in Section~\ref{results}.
\subsection{Datasets}
\label{datasets}
We conduct an extensive evaluation of session recommendation with PISA using two large-scale datasets from the music~domain:

\begin{itemize}
\item Last.fm \cite{schedl_icmr16}: This public dataset consists of over a billion time-stamped listening events from 
120k users of the music website Last.fm, encompassing 3M songs. 
We have filtered this dataset to include only the most recent year of consumption history in order for each user and each of the 15.7k songs 
to be associated with at least 1k and 1.5k listening actions respectively. 
\item Deezer: Our proprietary dataset contains over 700 million time-stamped listening events collected from 3.4M 
French users of the music streaming service Deezer. A listening event is defined as a user streaming a given track for at least 30 seconds, a threshold widely used in the industry for remuneration purposes. It includes 50k songs, among the most popular ones on the service. All events occurred between March and~August~2022.
\end{itemize}
We follow the methodology of Hansen et al.~\cite{hansen_recsys20} to group listening events into sessions, requiring at least 20 minutes of inactivity between each. In line with Section~\ref{s2}, we limit each session to the first $K=10$ songs.
We focus on users having at least 50 or 250 sessions for Last.fm and Deezer, respectively, and create sequences of 21 or 31 sessions using a sliding window over each user's session history with a step of five and twenty sessions respectively.
Our final datasets include about 465k sessions for Last.fm 
and 2.1M sessions for Deezer.

\subsection{Task and Evaluation Metrics}

\subsubsection{Task}
For both datasets, we form test sets from the last 10 sequences of each user, and validation sets from the preceding 5 ones. We observe the first $L=20$ or $L=30$ sessions of each sequence for Last.fm and Deezer, respectively. The $21^{\text{th}}$ or $31^{\text{th}}$ session is masked and set as the target for prediction. We evaluate the ability of PISA and baseline models to correctly retrieve the $K=10$ songs of the masked last session of each sequence, based on the preceding ones. In our experiments, each model must recommend lists of 10 songs, ranked by predicted relevance scores.

\subsubsection{Evaluation}
We consider nine different evaluation metrics for the above task. Firstly, we analyze two global accuracy metrics:
\begin{itemize}
    \item Recall: this score indicates which percentage of the ground truth 10 songs of each target session appear among the 10 songs recommended by each model. We report the average Recall score among all test sessions.
    \item Normalized Discounted Cumulative Gain (NDCG): this score acts as a
measure of ranking quality. Computed as in Equation (2) of Wang et al. \cite{wang2013theoretical}, it increases when ground truth songs are placed higher in the ranked list of 10 recommended songs. We report the average NDCG among all test sessions.

\end{itemize}
We also look deeper into the session compositions from a repetition and exploration perspective~\cite{li_infosys23}. Specifically, we compute:
\begin{itemize}
    \item $\text{Recall}^{\text{Rep}}$ and $\text{NDCG}^{\text{Rep}}$: two variants of Recall and NDCG, respectively, but computed only on \textit{repeated} songs of each ground truth target session, i.e., songs that have been listened to at least once in the previous sessions in user's history.
    \item $\text{Recall}^{\text{Exp}}$ and $\text{NDCG}^{\text{Exp}}$: two other variants of Recall and NDCG, respectively, but computed only on \textit{explored} songs of each ground truth target session, i.e., songs that have \textit{not} been listened to by users in their previous sessions.
\end{itemize}
For these metric pairs, we report average scores among all test sessions having at least one repeated or explored song,~respectively.
Finally, we gather insights from \text{beyond-accuracy}~\cite{kaminskas_tiis17} metrics:

\begin{itemize}
    \item Repetition Bias (RepRatio and RepBias): we verify whether recommendations lean towards repetition or exploration. RepRatio~\cite{li_infosys23} measures the average proportion of repeated songs in each recommended list of 10 songs. We compare it to RepRatio-GT, the ground truth average proportion in test sessions, by computing $\text{RepBias} = \text{RepRatio}~-~\text{RepRatio-GT}$. A positive (resp., negative) RepBias indicates a bias towards repetitive (resp., exploratory) recommendations.
   \item Popularity Bias: we compute the average intra-session Median Rank (MR) of recommended songs. The rank of a song is the number of users who have listened to it. Lower MR scores indicate the recommendation of less popular songs.

\end{itemize}

\subsection{Models}

\subsubsection{Two variants of PISA} 
We evaluate two variants of PISA, both based on the architecture outlined in Section~\ref{section_pepsi} but differing in the negative sampling techniques they use when evaluating the loss of Equation~\eqref{loss} during training. The first variant, denoted PISA-U in the following, uniformly samples the 10 songs appearing in each negative set $o^{(u)}_{l+1}$ from each unlistened song set $\mathcal{V} \setminus s^{(u)}_{l+1}$. The second variant, denoted PISA-P, alternatively employs a popularity-based negative sampling. 
In PISA-P, each song $v \in \mathcal{V} \setminus s^{(u)}_{l+1}$ is included in $o^{(u)}_{l+1}$ with a probability proportional to $f(v)^{\beta}$, where $f(v)$ represents the number of users who have listened to $v$ in the dataset, and $\beta = 1/2$ in our experiments. Consequently, popular songs are more likely to appear among negative samples.

\subsubsection{Baselines}
We compare PISA-U and PISA-P to ten baseline methods that cover a broad spectrum of NBR techniques. Firstly, we evaluate G-Top, a popularity baseline recommending the top-10 most listened to songs in the dataset to all users. We also assess P-Top, which recommends each user's top-10 most listened to songs. We note that this simple P-Top method is considered a quite strong NBR baseline \cite{li_infosys23, ariannezhad_sigir22}, despite only recommending repetitions.
We also present results from SASRec~\cite{kang_icdm18} to evaluate the direct use of a Transformer sequential recommender system without explicit modeling of repetitive behaviors as well as results from RepeatNet\footnote{RepeatNet was designed for "next-item" sequential recommendation. We use BCEWithLogitsLoss instead of negative log likelihood loss in order to adapt the model to the "next-basket" sequential recommendation setting.}~\cite{ren_aaai19}, a representative approach explicitly taking into account repetitive behaviors for sequential recommendation. 
Additionally, we highlight three repeat-aware NBR methods among the ones\footnote{Although ReCANet~\cite{ariannezhad_sigir22} and BRL~\cite{yu_ijcai23} are undoubtedly relevant baselines, we excluded them due to their demanding training times : over 5 and 10 hours per epoch on an NVIDIA RTX A5000 GPU, with about 20 epochs needed for convergence. Our experiments will also require averaging results over 5 runs, further extending the~total~time.} presented in Section~\ref{repeatNBRbaselines}: \text{TIFU-KNN} \cite{hu_sigir20}, UP-CF@r \cite{faggioli_umap20}, and Sets2Sets \cite{hu_sigkdd19}. Finally, in the music domain, we evaluate the non-repeat-aware CoSeRNN~\cite{hansen_recsys20} for listening session recommendation, as well as the two repeat-aware models using ACT-R~\cite{reiterhaas_recsys21,marta_recsys23}. We denote by ACT-R-Repeat the model of Reiter-Haas et al.~\cite{reiterhaas_recsys21}, which only recommends repeated songs. We denote by ACT-R-BPR the extension of ACT-R-Repeat by Moscati et al.~\cite{marta_recsys23}, which recommends repeated and new songs by combining ACT-R with a Bayesian
Personalized Ranking (BPR)~\cite{rendle_auai09} recommender system. 

\subsubsection{Implementation Details}
\label{implem-details}
PISA-U and PISA-P use pre-trained song embedding matrices $\textbf{M}$ for both datasets. 
For Last.fm, we compute song embedding vectors using a metric learning approach involving a triplet loss~\cite{weinberger2009distance} to bring closer vectors of songs listened to by the same users.
For Deezer, we use song embedding vectors provided by the Deezer service and obtained from the Singular Value Decomposition (SVD) of a mutual information matrix measuring song co-occurrences in Deezer playlists.
We train PISA-P, PISA-U and all neural network baselines for a maximum of 100 epochs using the Adam optimizer~\cite{kingma_iclr15} and batch sizes of 512. We set $d =$ 128 for all embedding-based models, $\alpha = 1/2$ for the BL module of all ACT-R models and $B =$ 2, $H =$ 2 for all Transformer-based models. Also, we recall that $K=10$, $\beta = 1/2$, $L=20$ for Last.fm, and $L=30$ for Deezer. We selected all other hyperparameters via a grid search on validation sets. For brevity, we report optimal values of all models in our public GitHub repository (see Section~\ref{implem}). Most notably, we tested learning rate values in the set \{0.0002, 0.0005, 0.00075, 0.001\} and $\lambda$ values in \{0.0, 0.3, 0.5, ~0.8, ~0.9, ~1.0\} .

\subsection{Open-Source Code and Data Release}
\label{implem}

Alongside this paper, we publicly release our TensorFlow implementation of PISA on GitHub\footnote{\label{sourcecode}https://github.com/deezer/recsys24-pisa} as well as our entire experimentation pipeline. Our aim is to ensure full reproducibility of our results and facilitate the future usage of our proposed PISA system.

Furthermore, we also release an anonymized version of our Deezer proprietary dataset on Zenodo, which is accessible from our GitHub\textsuperscript{\ref{sourcecode}}. This dataset is provided in its raw form, i.e., prior to the preprocessing operations of Section~\ref{datasets}, and includes our pre-trained song embedding vectors. By making these industrial resources available, we hope to support the scientific community and encourage future research in the field.

\subsection{Results and Discussion}
\label{results}

\begin{table*}[t]
  \caption{Listening session recommendation on Last.fm and Deezer, using PISA and other baselines. All models recommend ranked lists of 10 songs based on their predicted relevance. Scores are computed on test sets and averaged over five runs. All metrics should be maximized, except MR (minimized), RepRatio (close to the ground truth RepRatio-GT), and RepBias (close to 0). Bold and underlined numbers correspond to the best and second-best performance for each metric, respectively.}
  \label{acc_results}
  \resizebox{\textwidth}{!}{
  \begin{tabular}{c|c|c||cc|cc|cc|ccc}
        \toprule
        \multirow{2}{*}{\textbf{Dataset}} &  \multirow{2}{*}{\textbf{Model}} & {\textbf{Repetition}} & \multicolumn{2}{c}{\textbf{Global Metrics}} & \multicolumn{2}{c}{\textbf{Repetition-Focused Metrics}} & \multicolumn{2}{c}{\textbf{Exploration-Focused Metrics}} & \multicolumn{3}{c}{\textbf{Beyond-Accuracy Metrics}} \\
        \cline{4-12}
         &  & \textbf{Modeling} & {NDCG (in \%)} & {Recall (in \%)} & {$\text{NDCG}^{\text{Rep}} \text{ (in \%)}$} & {$\text{Recall}^{\text{Rep}}\text{ (in \%)}$} & {$\text{NDCG}^{\text{Exp}}\text{ (in \%)}$} & {$\text{Recall}^{\text{Exp}}\text{ (in \%)}$} & RepRatio (in \%) & RepBias & MR \\
                \midrule
        \midrule
        \multirow{12}{*}{\shortstack{\textbf{Last.fm}\\ \\ \\RepRatio-GT = $72.37\%$}} &  {G-Top} & {$\times$} & {$1.36 \pm 0.06$} & {$1.29 \pm 0.05$} & {$1.50 \pm 0.07$} & {$1.61 \pm 0.08$} & {$0.62 \pm 0.03$} & {$0.71 \pm 0.03$} & {$21.22 \pm 0.22$} & {$-51.15 \pm 0.22$} & {$40.20 \pm 0.00$} \\
        {} & {SASRec} & {$\times$}  & {5.36 $\pm$ 0.11} & {5.15 $\pm$ 0.11} & {5.65 $\pm$ 0.11} & {6.08 $\pm$ 0.12} & {1.90 $\pm$ 0.06} & {2.21 $\pm$ 0.07} & {48.38 $\pm$ 0.27} & {-23.99 $\pm$ 0.27} & {18.71 $\pm$ 0.08} \\
        {} & {CoSeRNN} & {$\times$}  & {6.99 $\pm$ 0.04} & {6.65 $\pm$ 0.03} & {6.55 $\pm$ 0.08} & {6.87 $\pm$ 0.09} & {3.22 $\pm$ 0.10} & {3.82 $\pm$ 0.09} & {42.03 $\pm$ 0.25} & {-30.32 $\pm$ 0.25} & {\underline{\textit{7.44 $\pm$ 0.03}}} \\
        {}  & {P-Top} & \checkmark & {8.34 $\pm$ 0.09} & {8.10 $\pm$ 0.09} & {11.32 $\pm$ 0.19} & {11.91 $\pm$ 0.21} & {0.00 $\pm$ 0.00} & {0.00 $\pm$ 0.00} & {100.00 $\pm$ 0.00} & {27.63 $\pm$ 0.00} & {10.75 $\pm$ 0.04} \\
        {} &  {UP-CF@r} & \checkmark & {7.31 $\pm$ 0.07} & {7.06 $\pm$ 0.07} & {9.70 $\pm$ 0.13} & {10.00 $\pm$ 0.13} & {0.02 $\pm$ 0.00} & {0.07 $\pm$ 0.01} & {97.49 $\pm$ 0.05} & {25.12 $\pm$ 0.05} & {14.89 $\pm$ 0.11} \\       
        {} & {Sets2Sets} & \checkmark & {6.90 $\pm$ 0.12} & {6.75 $\pm$ 0.12} & {7.55 $\pm$ 0.17} & {8.18 $\pm$ 0.19} & {2.11 $\pm$ 0.07} & {2.64 $\pm$ 0.05} & \textbf{77.13 $\pm$ 0.27} & \textbf{4.76 $\pm$ 0.27} & {11.14 $\pm$ 0.07} \\
        {} & {TIFU-KNN} & \checkmark & {9.05 $\pm$ 0.13} & {8.71 $\pm$ 0.13} & {\underline{\textit{12.33 $\pm$ 0.22}}} & {\underline{\textit{12.88 $\pm$ 0.22}}} & {0.07 $\pm$ 0.01} & {0.13 $\pm$ 0.02} & {99.04 $\pm$ 0.06} & {26.67 $\pm$ 0.06} &  {15.95 $\pm$ 0.06} \\
        {} & {RepeatNet} & \checkmark & {5.02 $\pm$ 0.08} & {4.97 $\pm$ 0.07} & {5.84 $\pm$ 0.11} & {6.14 $\pm$ 0.10} & {1.07 $\pm$ 0.02} & {1.38 $\pm$ 0.02} &  {46.41 $\pm$ 0.19} &  {-25.96 $\pm$ 0.19} &  {\textbf{7.22 $\pm$ 0.03}} \\
        {} &  {ACT-R-Repeat} & \checkmark & {9.18 $\pm$ 0.19} & {9.12 $\pm$ 0.19} & {\textbf{13.94 $\pm$ 0.18}} & {\textbf{15.75 $\pm$ 0.19}} & {0.00 $\pm$ 0.00} & {0.00 $\pm$ 0.00} & {100.00 $\pm$ 0.00} & {27.63 $\pm$ 0.00} & {8.29 $\pm$ 0.05} \\
        {} & {ACT-R-BPR} & \checkmark & {3.07 $\pm$ 0.03} & {3.02 $\pm$ 0.03} & {4.11 $\pm$ 0.05} & {4.49 $\pm$ 0.06} & {0.38 $\pm$ 0.01} & {0.60 $\pm$ 0.02} & {\underline{\textit{79.65 $\pm$ 0.18}}} & \underline{\textit{7.28 $\pm$ 0.18}} & {7.85 $\pm$ 0.03} \\
       \cline{2-12}
        {} & {PISA-U (ours)} & \checkmark & \underline{\textit{{12.09 $\pm$ 0.13}}} & \underline{\textit{{11.59 $\pm$ 0.13}}} & {11.51 $\pm$ 0.15} & {12.24 $\pm$ 0.13} & \underline{\textit{{5.45 $\pm$ 0.06}}} & \underline{\textit{{6.09 $\pm$ 0.06}}} & {61.23 $\pm$ 0.19} & {-11.14 $\pm$ 0.19} & {9.35 $\pm$ 0.06} \\
        {} & {PISA-P (ours)} & \checkmark  & {\textbf{12.16 $\pm$ 0.16}} & {\textbf{11.77 $\pm$ 0.13}} & {11.49 $\pm$ 0.16} & {12.22 $\pm$ 0.15} & {\textbf{5.50 $\pm$ 0.08}} & {\textbf{6.16 $\pm$ 0.10}} & {61.63 $\pm$ 0.09} & {-10.74 $\pm$ 0.09} & {8.24 $\pm$ 0.05} \\
        \midrule
        \midrule
        \multirow{12}{*}{\shortstack{\textbf{Deezer} \\ \\ \\ RepRatio-GT = 89.10\%}} & {G-Top} & {$\times$} & {4.40 $\pm$ 0.08} & {3.90 $\pm$ 0.07} & {4.45 $\pm$ 0.09} & {4.29 $\pm$ 0.09} & {2.05 $\pm$ 0.05} & {2.52 $\pm$ 0.06} & {67.19 $\pm$ 0.51} & {-21.91 $\pm$ 0.51} & {123.57 $\pm$ 0.00} \\
        {} & {SASRec} & {$\times$} & {6.05 $\pm$ 0.10} & {5.89 $\pm$ 0.09} & {6.47 $\pm$ 0.10} & {6.71 $\pm$ 0.09} & {0.60 $\pm$ 0.04} & {1.12 $\pm$ 0.08} & {85.65 $\pm$ 0.25} & {-3.45 $\pm$ 0.25} & {67.56 $\pm$ 0.10} \\ 
        {} & {CoSeRNN} & {$\times$} & {2.57 $\pm$ 0.06} & {2.46 $\pm$ 0.06} & {2.69 $\pm$ 0.07} & {2.73 $\pm$ 0.06} & {0.44 $\pm$ 0.04} & {0.76 $\pm$ 0.04} & {71.48 $\pm$ 0.31} & {-17.62 $\pm$ 0.31} & {49.14 $\pm$ 0.21} \\   
        {} &  {P-Top} & \checkmark & {7.83 $\pm$ 0.18} & {7.51 $\pm$ 0.16} & {8.37 $\pm$ 0.18} & {8.39 $\pm$ 0.16} & {0.00 $\pm$ 0.00} & {0.00 $\pm$ 0.00} & {100 $\pm$ 0.00} & {10.9 $\pm$ 0.00} & {45.33 $\pm$ 0.38} \\      
        {} &  {UP-CF@r} & \checkmark & {9.88 $\pm$ 0.23} & {9.35 $\pm$ 0.21} & {10.62 $\pm$ 0.23} & {10.54 $\pm$ 0.21} & {0.11 $\pm$ 0.01} & {0.26 $\pm$ 0.03} & {96.87 $\pm$ 0.09} & {7.77 $\pm$ 0.09} & {66.21 $\pm$ 0.16} \\
        {} &  {Sets2Sets} & \checkmark & {7.84 $\pm$ 0.21} & {7.40 $\pm$ 0.18} & {8.03 $\pm$ 0.24} & {7.97 $\pm$ 0.21} & {1.43 $\pm$ 0.10} & {2.62 $\pm$ 0.19} & \underline{\textit{91.29 $\pm$ 0.10}} & \underline{\textit{2.19 $\pm$ 0.10}} & {31.94 $\pm$ 0.25} \\       
        
        {} & {TIFU-KNN} & \checkmark & {10.22 $\pm$ 0.26} & \underline{\textit{{9.64 $\pm$ 0.22}}} & {\underline{\textit{11.07 $\pm$ 0.27}}} & {\underline{\textit{10.95 $\pm$ 0.24}}} & {0.22 $\pm$ 0.01} & {0.51 $\pm$ 0.03} & {94.80 $\pm$ 0.10} & {5.70 $\pm$ 0.10} & {88.33 $\pm$ 0.33} \\
        {} & {RepeatNet} & \checkmark & {1.25 $\pm$ 0.02} & {1.26 $\pm$ 0.01} & {1.25 $\pm$ 0.02} & {1.33 $\pm$ 0.01} & {0.32 $\pm$ 0.02} & {0.53 $\pm$ 0.04} &  {31.81 $\pm$ 0.31} &  {-57.29 $\pm$ 0.31} &  {\textbf{17.65 $\pm$ 0.11}} \\
        
        {} &  {ACT-R-Repeat} & \checkmark & {7.93 $\pm$ 0.17} & {7.95 $\pm$ 0.17} & {8.88 $\pm$ 0.16} & {9.59 $\pm$ 0.15} & {0.00 $\pm$ 0.00} & {0.00 $\pm$ 0.00} & {100 $\pm$ 0.00} & {10.90 $\pm$ 0.00} & {\underline{\textit{31.45 $\pm$ 0.12}}} \\
        {} &  {ACT-R-BPR} & \checkmark & {2.38 $\pm$ 0.01} & {2.36 $\pm$ 0.01} & {2.51 $\pm$ 0.02} & {2.66 $\pm$ 0.02} & {0.42 $\pm$ 0.03} & {0.78 $\pm$ 0.07} & {80.10 $\pm$ 0.30} & {-9.00 $\pm$ 0.30} & {38.13 $\pm$ 0.24} \\      
        \cline{2-12}     
        {} &  {PISA-U (ours)} & \checkmark & \underline{\textit{{10.27 $\pm$ 0.09}}} & {9.54 $\pm$ 0.12} & {10.46 $\pm$ 0.12} & {10.49 $\pm$ 0.12} & \underline{\textit{{2.06 $\pm$ 0.05}}} & {\underline{\textit{3.11 $\pm$ 0.05}}} & {\textbf{88.27 $\pm$ 0.10}} & {\textbf{-0.83 $\pm$ 0.10}} & {55.95 $\pm$ 0.26} \\    
        {} & {PISA-P (ours)} & \checkmark & {\textbf{11.20 $\pm$ 0.13}} & {\textbf{10.40 $\pm$ 0.14}} & {\textbf{11.08 $\pm$ 0.14}} & {\textbf{11.07 $\pm$ 0.07}} & {\textbf{3.13 $\pm$ 0.07}} & {\textbf{4.54 $\pm$ 0.12}} & {85.16 $\pm$ 0.08} & {-3.94 $\pm$ 0.08} & {39.45 $\pm$ 0.23} \\
        \bottomrule
      \end{tabular}
    }
\end{table*}

Table~\ref{acc_results} presents all test set results, averaged over five runs with standard deviations. Overall, PISA shows competitive performances. We obtain the highest scores in 10/12 NDCG and Recall metrics (e.g., a top 12.16\% NDCG for PISA-P on Last.fm), which we discuss~below.
 
\subsubsection{On the Importance of Repetition Modeling}
Firstly, our experiments demonstrate that repeat-aware methods generally outperform the non-repeat-aware ones across most accuracy metrics. This result confirms the critical importance of modeling repetitive patterns for effective listening session recommendation on real-world music streaming services, where users frequently relisten to songs. We note that the ground truth average proportion of repeated songs in test sessions to retrieve, i.e., RepRatio-GT, is relatively high in both datasets: 72.37\% for Last.fm and 89.10\% for Deezer.

\subsubsection{PISA vs Other Repeat-Aware Methods}

Accounting for repetitions is crucial, but the method employed to achieve this goal is equally important. PISA-P achieves the best performance across all four global accuracy metrics, followed by PISA-U, which ranks second in three of these metrics. Further analysis of the composition of recommended sessions shows that, while repeat-aware baselines perform well on repetition-focused metrics, their scores severely decline on exploration-focused ones (e.g., TIFU-KNN reaches a second-best 12.33\% $\text{NDCG}^{\text{Rep}}$ on Last.fm, but a low 0.07\% $\text{NDCG}^{\text{Exp}}$). In contrast, PISA-U and PISA-P not only predict repeated songs with comparable effectiveness, but also significantly outperform baselines in recommending songs that users have not yet listened to (e.g., PISA-P simultaneously reaches a top 11.08\% $\text{NDCG}^{\text{Rep}}$ and top 3.13\% $\text{NDCG}^{\text{Exp}}$ on Deezer).
This is an important result, as enhanced exploration helps users discover new territories within the vast catalog of music streaming services. We postulate that leveraging ACT-R for session and user embedding not only allows PISA to capture repetitive patterns but also to focus on songs that effectively represent users' musical tastes, potentially improving exploration. Note that, even though repeated songs affect the position of user embedding vectors, PISA models will still recommend a new song if it demonstrates high predicted relevance in the embedding space, as determined in Section~\ref{prediction}. This approach contrasts with other baselines (P-Top, ACT-R-Repeat) that recommend sessions composed solely of repeated songs by design. 
Finally, we found that sequential models (Sets2sets, PISA, and even the non-repeat-aware SASRec and CoseRNN) outperform non-sequential models for exploration, confirming the benefits of dynamic preference modeling.

\subsubsection{PISA vs Other ACT-R Methods}
PISA outperforms ACT-R baselines from Table~\ref{acc_results} in most NDCG and Recall metrics. Admittedly, previous work primarily focused on relistening prediction (for ACT-R-Repeat) and explainability (for ACT-R-BPR), rather than performance\footnote{Although we note that ACT-R-Repeat performs rather well in repetition, likely because it only considers the smaller candidate set of repeated songs for recommendations.}. Notwithstanding, unlike these methods, PISA's Transformer harnesses the dynamic dimension of preferences. Our results confirm that modeling dynamic patterns is critical in our problem. This conclusion is corroborated by the performance of sequential models like SASRec and CoSeRNN, which achieve comparable or superior results to ACT-R baselines without modeling~repetitions.

\subsubsection{On Repetition and Popularity Biases}
Beyond performance, our analysis of RepRatio and RepBias shows that non-repeat-aware approaches are biased towards exploration when recommending sessions, i.e., they underestimate how often users relisten to songs. Meanwhile, repeat-aware approaches based on frequency (P-Top) or nearest neighbors  (UP-CF@r, TIFU-KNN) overestimate repetitions, often by over 20 percentage points for Last.fm. In contrast, Set2Sets and the ACT-R-based ACT-R-BPR, PISA-U, and PISA-P achieve a better balance between these consumption modes, more closely matching ground truth repetition ratios (e.g., with a top -0.83\% RepBias for PISA-P on Deezer). This result underscores their effectiveness in aligning recommendations with actual user behavior regarding repetition and exploration.
Besides, ACT-R-BPR and PISA-P are less prone to popularity bias than most baselines, which is often viewed as a desirable property \cite{schedl2018current}. They are among the methods with quite low MR scores for both datasets, with a significant improvement over mainstream baselines such as G-Top.

\subsubsection{On PISA Components}
We noticed that attention weights for the partial matching component $\text{P}^{(u)}_v$ consistently converged to zero\footnote{On the contrary, BL weights were consistently the highest in all four PISA models.} in our experiments. Hence, PISA models overlooked this component, likely because our pre-trained song embeddings (derived from collaborative filtering -- see Section~\ref{implem-details}), capture similar information to the spreading component instead of content-based information\footnote{Although PISA can automatically learn to diminish the relative importance of irrelevant ACT-R components for specific tasks or datasets, our public implementation of PISA will allow future users to manually exclude any ACT-R component before training. This feature allows future users to train an ablated version of PISA by removing ACT-R components that they know will not be relevant to their specific applications.}.
Furthermore, the choice of negative sampling method influences results. Substituting uniform negative sampling (PISA-U) with popularity sampling (PISA-P) improves the effectiveness of PISA, including its exploration capabilities. It also helps reduce popularity biases (e.g., from 55.95\% to 39.45\% MR on Music Streaming). These findings are consistent with prior research, which has discussed the advantages of popularity sampling over uniform sampling for better personalization by training models on more relevant negative items \cite{tran_sigir19, pellegrini_recsys22}. Based on our results, we suggest that future PISA users opt for the PISA-P variant.

\section{Conclusion and Future Work}
\label{conclusion}

In conclusion, the primary contribution of our work is the development of PISA, a novel system tailored for repeat-aware and sequential listening session recommendation. We have integrated the ACT-R cognitive architecture within a Transformer-based model, enabling us to jointly capture dynamic and repetitive patterns from listening session sequences. Our experiments have confirmed the empirical relevance of PISA for warm-start scenarios. Furthermore, we have released our dataset of listening sessions from the Deezer service to foster further research. Overall, our findings underscore the critical importance of modeling repetitive patterns for effective sequential listening session recommendation. 
They also open up interesting avenues for future research. We intend to explore the  addition of other psychological modules. For example, PISA does not explicitly account for curiosity, or interest for new items, a limitation of models based on memory i.e., the past. 
Modeling this dynamic would be beneficial.

\bibliographystyle{ACM-Reference-Format}
\bibliography{references}

\end{document}